\documentclass[pra,twocolumn,floatfix,showpacs,superscriptaddress]{revtex4}

\usepackage{amsmath,amsfonts,amssymb,bm}
\usepackage{dcolumn}
\usepackage[final]{graphicx}
\usepackage{bm}
\usepackage{comment}

\includecomment{pdffigure}
\begin{document}

\bibliographystyle{apsrev}	

\title{Ab--initio approach for gap plasmonics}

\author{Ulrich Hohenester}
\email{ulrich.hohenester@uni-graz.at}

\affiliation{Institute of Physics,
  University of Graz, Universit\"atsplatz 5, 8010 Graz, Austria}
\affiliation{
  Humboldt-Universit\"at zu Berlin,
  Physics Department and IRIS Adlershof,
  Zum Gro\ss en Windkanal 6, 12489 Berlin, Germany}  
\author{Claudia Draxl}

\affiliation{
  Humboldt-Universit\"at zu Berlin,
  Physics Department and IRIS Adlershof,
  Zum Gro\ss en Windkanal 6, 12489 Berlin, Germany}

\date{September 28, 2016)}

\begin{abstract}
Gap plasmonics deals with the properties of surface plasmons in the narrow region between two metallic nanoparticles forming the gap.  For sub-nanometer gap distances electrons can tunnel between the nanoparticles leading to the emergence of novel charge transfer plasmons.  These are conveniently described within the quantum corrected model by introducing an artificial material with a tunnel conductivity inside the gap region.  Here we develop a methodology for computing such tunnel conductivities within the first-principles framework of density functional theory, and apply our approach to a jellium model representative for sodium.  We show that the frequency dependence of the tunnel conductivity at infrared and optical frequencies can be significantly more complicated than previously thought.
\end{abstract}


% 73.20.Mf Surface plasmons.
% 78.67.Bf Optical properties of nanocrystalline materials.
% 03.50.De Classical electrodynamics.

\maketitle

%%%%%%%%%%%%%%%%%%%%%%
%%%  INTRODUCTION  %%%
%%%%%%%%%%%%%%%%%%%%%%

\section{Introduction}

Plasmonics achieves light confinement at the nanoscale by binding light to coherent charge oscillations of metallic nanoparticles, so-called surface plasmons (SPs)~\cite{maier:07,stockman:11}.  Gap plasmonics deals with SPs in gap regions of coupled metallic nanoparticles where extremely high field enhancements can be achieved.  For gap distances in the sub-nanometer range electrons can tunnel through the gap region, leading to the emergence of new charge transfer plasmons (CTPs)~\cite{esteban:12,esteban:15,zhu:16} which have been observed optically~\cite{savage:12,zhu:14} and in electron energy loss spectroscopy (EELS)~\cite{duan:12,scholl:13}.  Tunneling through larger gap regions has been demonstrated in molecular tunnel junctions~\cite{tan:14,cha:14,benz:15,knebl:16}.

The reconciliation of electrodynamic and quantum effects provides a serious challenge from the theoretical side.  For small nanoparticles with dimensions of a few nanometers, consisting of several tens to hundreds of atoms, one can directly employ atomistic simulation approaches, such as time dependent density functional theory (TDDFT)~\cite{esteban:12,zhang:14,varas:15,barbry:15,kulkarni:15,xiang:16}.  However, for larger nanoparticles with dimensions in the tens to hundreds of nanometer range such approach is doomed to failure and one must resort to effective models, such as the quantum corrected model (QCM)~\cite{esteban:12,esteban:15,hohenester.prb:15,zhu:16}:  Here, one introduces an artificial material inside the gap region whose conductivity $\sigma_t$ mimics electron tunneling.  In their original work, Esteban et al.~\cite{esteban:12,esteban:15} proposed a Drude-type expression for $\sigma_t$ that interpolates between the bulk properties of the metal at small gap distances and an exponential tunnel decay at large gap distances.  Once $\sigma_t$ of the artificial material is chosen, it can be incorporated into standard simulation solvers for electrodynamic problems.

The question of how to combine electrodynamic and quantum descriptions is also encountered for SPs and nonlocality, which typically plays an important role for small nanoparticles or nanoparticles with sharp features:  Electron pressure leads to a spill-out of the electrons, resulting in a blue-shift of the SP resonances with respect to local dielectric descriptions~\cite{ciraci:12,scholl:12}.  To account for nonlocal effects, one can add an artificial material layer around the particle~\cite{luo:13} or employ a hydrodynamic model~\cite{david:11,ciraci:12,mortensen:14,david:14} where all nonlocal effects are lumped into a few effective hydrodynamic parameters.  It has been demonstrated recently~\cite{toscano:15} that nonlocal plasmonics can be formulated self-consistently within a combined electrodynamic and hydrodynamic framework, where the pertinent parameters are obtained independently from density functional theory (DFT) simulations.  

In this paper we study gap plasmonics within the first principles approach of DFT.  We start by developing in Sec.~\ref{sec:theory} a methodology for the computation of the gap conductivity $\sigma_t$ that improves upon the interpolating expression of Esteban et al.~\cite{esteban:12,esteban:15}.  We derive a general Kubo-type formula for $\sigma_t$ and show how to incorporate image charge effects that have been claimed to be of importance for gap plasmonics~\cite{esteban:12,esteban:15}.  In Sec.~\ref{sec:results} we apply our methodology to a jellium model representative for sodium and show that the frequency and gap-distance dependence of $\sigma_t$ can be significantly more complicated than predicted by the usual interpolating expression.  Results for electrodynamic simulations using the ab-initio and interpolating $\sigma_t$ expressions turn out to be in fair agreement.  Finally, in Sec.~\ref{sec:discussion} we discuss guidelines for obtaining more reliable model expressions for $\sigma_t$.

%%%%%%%%%%%%%%%%
%%%  Theory  %%%
%%%%%%%%%%%%%%%%

\section{Theory}\label{sec:theory}

\begin{figure}
\centerline{\includegraphics[width=\columnwidth]{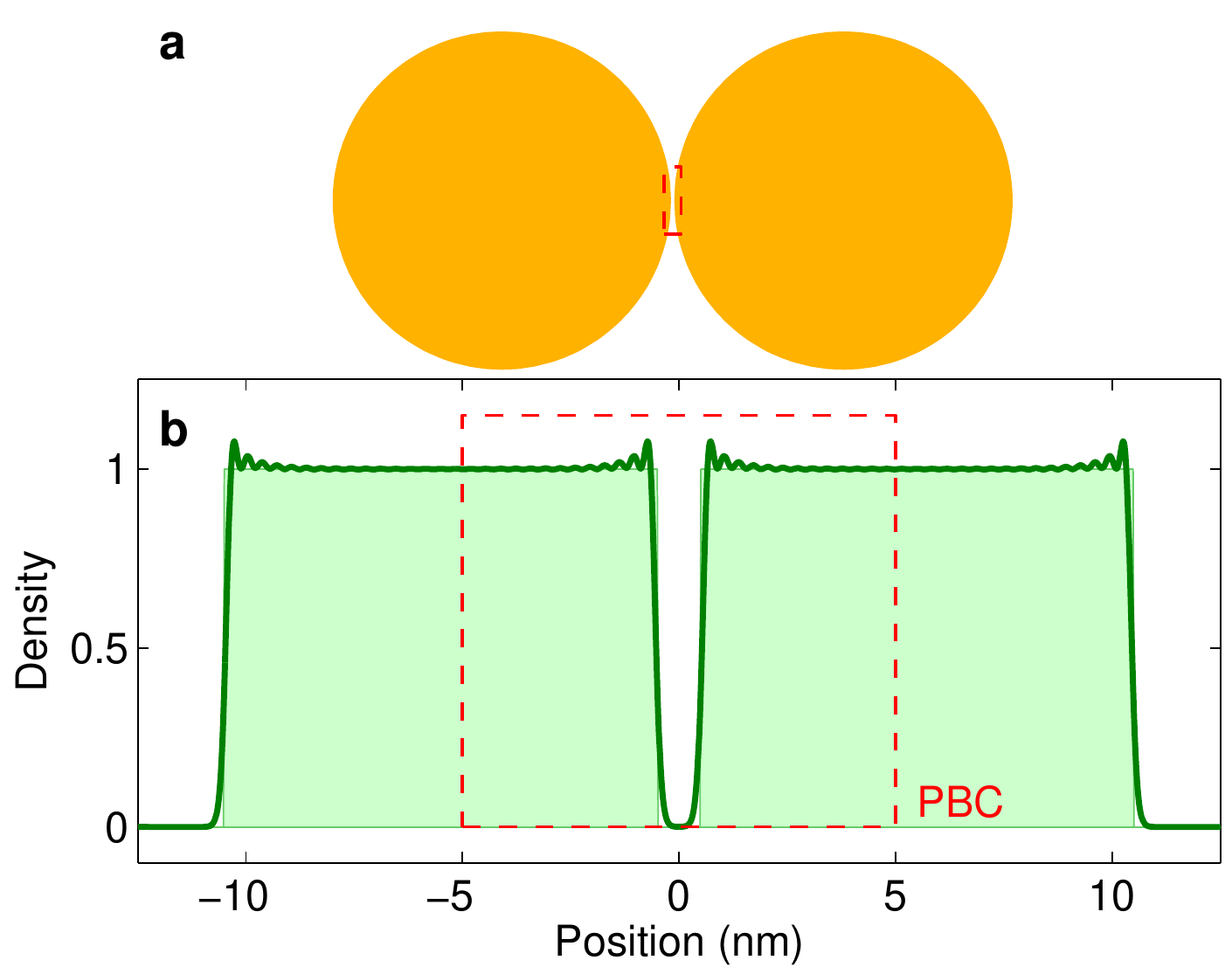}}
\caption{Gap plasmonics for (a) two coupled metallic nanoparticles is modeled by approximating the gap region by (b) the interspace between two metallic slabs and computing the tunnel conductivity $\sigma_t(\omega,\ell)$ within the framework of density functional theory (DFT) for various gap distances $\ell$.  In a consecutive step, $\sigma_t(\omega,\ell)$ is submitted to the quantum corrected model~\cite{esteban:12,esteban:15,hohenester.prb:15}.  Panel (b) reports results for a jellium model representative for sodium.  For a homogeneous external excitation one can use periodic boundary conditions (PBC), as discussed in the text.}
\end{figure}

\subsection{Quantum corrected model}  

We start by considering a plasmonic dimer with a narrow gap, as depicted in Fig. 1(a).  Upon external excitation, for instance by optical fields, a tunnel current flows between the two nanoparticles.  For sufficiently weak excitations, the induced tunnel current $J_t$ in the middle of the gap is proportional to the applied electric field $E$,
\begin{equation}
  J_t=\sigma_t E\,,
\end{equation}
where $\sigma_t$ is the tunnel conductivity defined in accordance to scanning tunneling microscopy (STM) \cite{blanco:06}.  For a metallic junction, Esteban and coworkers~\cite{esteban:12,esteban:15} suggested to use for the gap region a Drude-type gap permittivity
\begin{equation}\label{eq:drudegap}
  \varepsilon_g(\omega,\ell)=1-\frac{\omega_p^2}{\omega\bigl(\omega+i\gamma_g(\ell)\bigr)}\,,
\end{equation}
which is related to the tunnel conductivity through $\sigma_t=-i\omega[\varepsilon_g(\omega,\ell)-1]/(4\pi)$.  Here $\omega$ is the angular frequency, $\ell$ is the gap separation, $\omega_p$ is the bulk plasma frequency, and $\gamma_g(\ell)=\gamma_p\,e^{\ell/\ell_c}$ depends on the damping term $\gamma_p$ of the bulk dielectric function and a characteristic tunnel length scale $\ell_c$.  $\varepsilon_g(\omega,\ell)$ is constructed such that for zero gap distance the Drude dielectric function is recovered, whereas for large distances the conductivity decays exponentially $\sigma_t\sim e^{-\ell/\ell_c}$ in accordance to tunnel processes.  Eq.~\eqref{eq:drudegap} can be directly applied to free electron gases such as sodium, whereas for noble metals such as gold or silver one has to additionally consider $d$-band contributions~\cite{esteban:15,toscano:15}.  

To further motivate Eq.~\eqref{eq:drudegap}, Esteban et al.~\cite{esteban:12,esteban:15} additionally performed TDDFT simulations for small sodium spheres and demonstrated that the computed spectra agree well with those of classical electrodynamic simulations using $\varepsilon_g(\omega,\ell)$ for the gap permittivity.  They also obtained from tunneling theory within the WKB-approximation the static conductivity $\sigma_t(0,\ell)$, and showed that it indeed decays with the characteristic length scale $\ell_c$.

\subsection{First-principles approach} 

In the following we derive a methodology for calculating $\sigma_t$ within a first-principles DFT approach.  Our starting point is the Hamiltonian for a many electron system subject to an external perturbation, described by the vector potential $\bm A$ (minimal coupling)~\cite{mahan:81}
\begin{equation}\label{eq:ham}
  H=\sum_i\frac 12\left[\hat{\bm p}_i-q\bm A(\bm r_i,t)\right]^2+V\,.
\end{equation}
Here $\hat{\bm p}_i$ is the electron momentum operator, $q=-1$ is the electron charge, and $V$ the sum of external and electron-electron potentials.  We use atomic units $m=e=\hbar=1$ throughout.  In linear response and assuming a weak spatial variation of $\bm A$, the light-matter coupling can be approximately written in the form $H_{\rm op}=-q\sum_i\bm A(\bm r_i,t)\cdot\hat{\bm p}_i$.  The electric current $\bm J=q\sum_i\langle\hat{\bm v}_i\rangle$ depends on the electron velocity $\hat{\bm v}_i=\hat{\bm p}_i-q\bm A(\bm r_i,t)$ (canonical momentum), which can be rewritten within the framework of second quantization as~\cite{mahan:81}
\begin{equation}\label{eq:current}
  \bm J(\bm r,t)=q\left<\hat{\bm J}_1(\bm r,t)\right>-q^2n(r,t)\bm A(\bm r,t)\,.
\end{equation}
Here we have introduced the current operator $\hat{\bm J}_1=-\frac i2[\hat\psi^\dagger(\nabla\hat\psi)-(\nabla\hat\psi^\dagger)\hat\psi]$, the field operator $\hat\psi$ for electrons, and $\langle.\rangle$ denotes a suitable wavefunction and ensemble average.  The first and second term on the right hand side of Eq.~\eqref{eq:current} are usually referred to as the \textit{paramagnetic} and \textit{diamagnetic} current, respectively.  Within linear response theory the expectation value can be evaluated to obtain a Kubo-type formula~\cite{kubo:85}
\begin{equation}\label{eq:kubo}
  \bm J_1(\bm r,t)=i\lim_{t_0\to-\infty}\int_{t_0}^t\left<\left[\hat{\bm J}_1(\bm r,t),\hat H_{\rm op}(t')\right]\right>_0e^{\eta t'}\,dt'\,,
\end{equation}
where $\langle.\rangle_0$ denotes the expectation value for the unperturbed system.  We have assumed the usual adiabatic switching-on of the external perturbation with $\eta$ being a small positive quantity \cite{mahan:81}.  Eqs.~\eqref{eq:current} and \eqref{eq:kubo} allow quite generally to compute the tunnel conductivity for a plasmonic system within linear response.

\subsection{Jellium slabs}  

We shall now be more specific regarding the simulated system.  In accordance to Refs.~\cite{esteban:12,esteban:15}, we assume for nanoparticles with dimensions in the range of tens of nanometers that the curvature of the metallic nanoparticles is small in comparison to the gap distance, such that we can approximately model tunneling in the gap region by considering planar metal slabs with varying gap separations $\ell$, as shown in Fig.~1(b).  The total tunnel current is then obtained by integrating over the plane in the middle of the gap and parallel to the slab surface, $J_t=\oint J(\bm r)\,da$, where from now on we will only consider electric fields and currents oriented along the direction $x$ perpendicular to the slab surface.

The properties of the slab system are modeled within the DFT framework~\cite{dreizler:90}.  We interpret the Kohn-Sham energies $E_n$ (with corresponding wavefunctions $\phi_n$) as the band structure, as justified for electron-gas like metallic systems.  Image charge effects in the gap region~\cite{blanco:06} can be considered by using quasiparticle energies and wavefunctions computed within the GW approximation~\cite{aryasetiawan:98}, or, as we shall do in this work, by adopting the conceptually more simple weighted density approximation (WDA) using a non-isotropic exchange-correlation hole~\cite{gunnarson:79,gonzalez:00,gonzalez:03}.  With $E_n$, $\phi_n$ at hand, we can evaluate the current-current correlation function of Eq.~\eqref{eq:kubo} for a harmonic time dependence of the external perturbation, and finally arrive at
\begin{widetext}
\begin{equation}\label{eq:sigt}
  \sigma_t(\omega,\ell)=-\frac i\omega \sum_{mn}\frac{f_m-f_n}{\omega+E_m-E_n+i\eta}
  \int\bigl<m\bigr|\hat{J}_1(\bm r)\delta(x)\bigl|n\bigr>
   \bigl<n\bigr|\hat{J}_1(\bm r')\bigl|m\bigr>\,d\bm rd\bm r'+
   \frac i\omega\int n(\bm r)\delta(x)\,d\bm r\,.
\end{equation}
\end{widetext}
Here $f_m$ is the Fermi-Dirac distribution function for electrons (evaluated at either zero or finite temperature), and $n(\bm r)$ is the electron density.  Eq.~\eqref{eq:sigt} is the central result of this work.  It allows us to compute from first principles the tunnel conductivity, which consists of the \textit{paramagnetic} contribution given by the current-current correlation function and the \textit{diamagnetic} contribution proportional to $n(\bm r)$.

In this work we assume zero temperature throughout.  Quite generally, the non radiative decay of plasmons inevitably generates heat similar to the cases of nanoparticles~\cite{lereu:13} and thin films~\cite{passian:05}.  Therefore, for a sub-nm gap it might be expected that thermal effects cause some fluctuation in gap size and, depending upon the sensitivity of the tunnel current with 
the gap distance, introduce additional noise.  Such effects are not considered in this work.

\section{Results}\label{sec:results}

\subsection{The case of sodium slabs}  

In the following we submit Eq.~\eqref{eq:sigt} to a corrected jellium-type model representative for sodium, with a Wigner Seitz radius $r_s=3.93$ and an additional confinement potential of $1.15$ eV chosen to yield the correct work function~\cite{perdew:90}.  Our approach closely follows related work of Refs.~\cite{esteban:12,esteban:15,toscano:15}.  In our DFT simulations we first employ the local density approximation (LDA) and compute the ground state properties self consistently using the exchange correlation potential of Ref.~\cite{perdew:81}.  The self consistent LDA density is used as an input for WDA, and we solve the Kohn-Sham equations for the WDA exchange-correlation potential only once.  This approach is similar to the G$_0$W$_0$ approximation~\cite{aryasetiawan:98} or the WKB approach of Refs.~\cite{esteban:12,esteban:15} using a suitable image charge potential~\cite{pitarke:90}.

For the tunneling simulations we consider two jellium slabs, as depicted in Fig.~1(b), and Kohn-Sham energies and wavefunctions of the form $E_{n\bm k_\|}=\frac 12 k_\|^2+\epsilon_n^\perp$ and $\phi_{n\bm k_\|}(\bm r)=e^{i\bm k_\|\cdot\bm r}\varphi_n^\perp(x)$, respectively, where $\bm k_\|$ is the wavevector in the in-plane directions of the slabs and $\varphi_n^\perp(x)$ is a one-dimensional Kohn-Sham wavefunction computed on a real-space grid using finite differences (with typically a few thousand discretization points).  A technical detail that might be beneficial for realistic DFT simulations beyond the jellium model is that our initial Hamiltonian only depends on the vector potential $\bm A$, whose spatial variation can be neglected for optical wavelengths.  Thus, we can introduce in our simulation approach periodic boundary conditions, as schematically depicted in Fig.~1(b).  Within this approach we then simulate a super cell of slabs.  The advantage of periodic boundary conditions is that we do not have to introduce vacuum regions in the simulation domain which are usually computationally expensive.  Below we will demonstrate the validity of this approach, and will use periodic boundary conditions unless stated differently.

\begin{figure}
\centerline{\includegraphics[width=0.85\columnwidth]{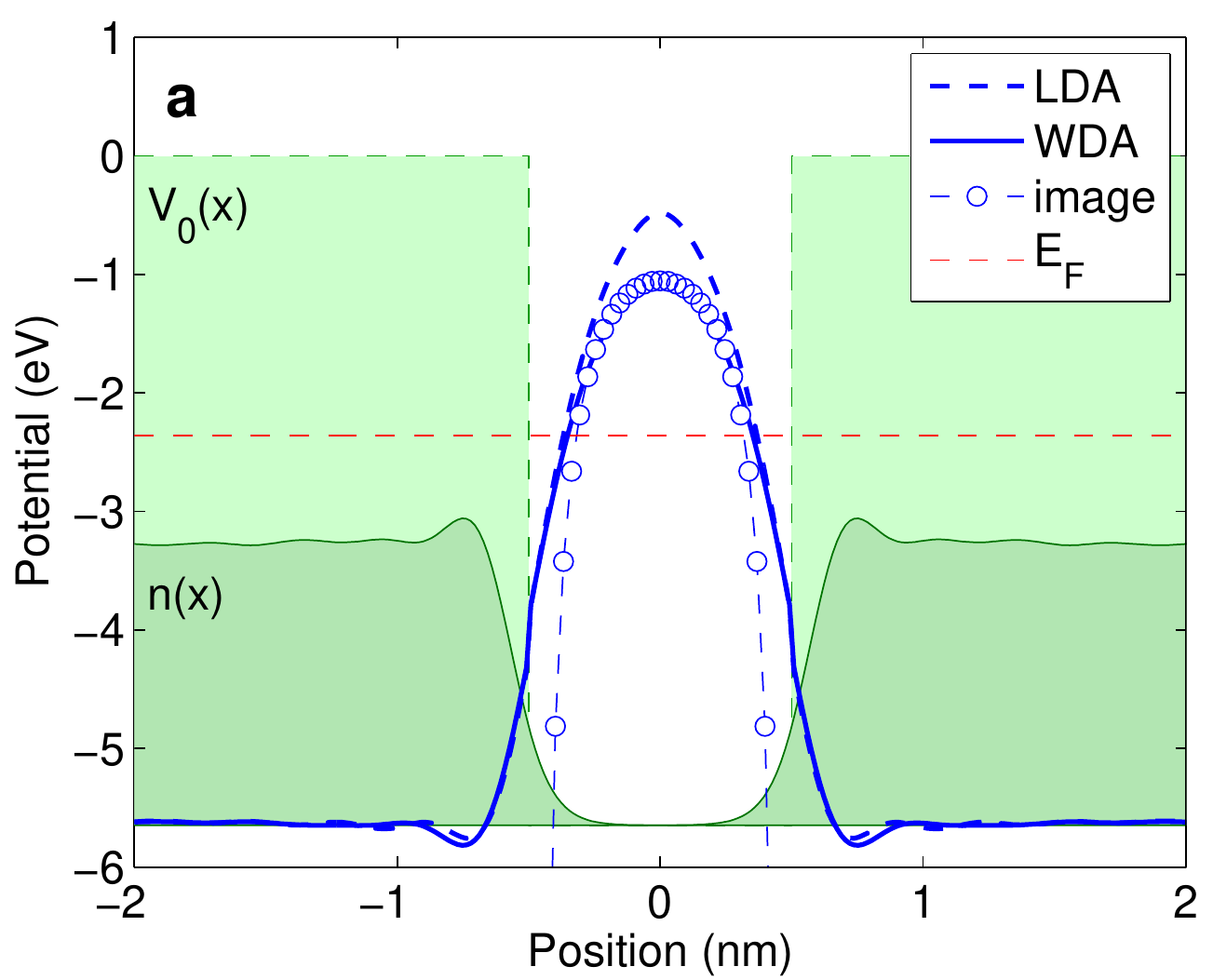}}
\centerline{\includegraphics[width=0.85\columnwidth]{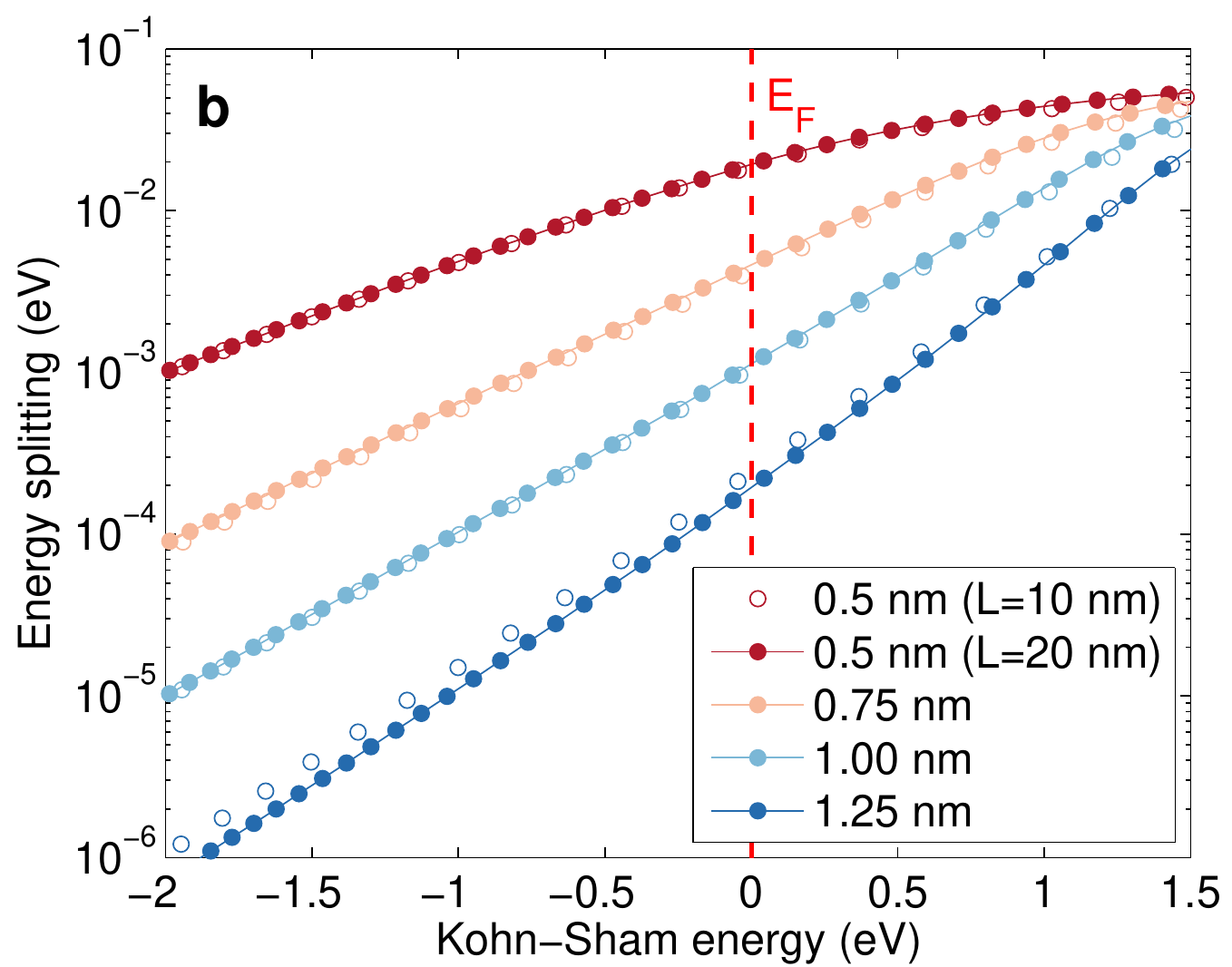}}
\caption{
(a) Density $n(x)$ (green shaded area), sum of exchange-correlation potential and external potential for corrected jellium model~\cite{perdew:90} as computed within the local (LDA, dashed line) and weighted (WDA, solid line) density approximations, and image charge potential (circles, image plane position of $0.5$ a.u.~\cite{gonzalez:03}) for two jellium slabs representative for sodium separated by 1 nm.  $E_F$ denotes the Fermi energy position.  (b) Energy splitting of WDA Kohn-Sham energies for selected gap distances and for slab widths of 20 nm (filled circles) and 10 nm (open circles).
}
\end{figure}

Fig.~2(a) shows for two jellium slabs separated by a gap distance of 1 nm the density profile $n(x)$ along with the sum of exchange-correlation and confinement potential computed within LDA and WDA, respectively.  As can be seen, in the gap region the WDA results (solid line) coincide perfectly well with the classical image charge potential~\cite{pitarke:90} (circles), whereas inside the jellium the LDA and WDA results are in very good agreement.  The small deviations can be attributed to the slightly different exchange correlation potentials of Refs.~\cite{perdew:81} and \cite{chacon:88} used in our implementations. 

For two uncoupled slabs the Kohn-Sham energies are double degenerate, one eigenvalue corresponding to the left and the other one to the right slab.  When the two slabs become coupled through tunneling, the eigenstates split into the usual bonding and antibonding states with an energy difference given by the tunnel coupling.  Fig.~2(b) shows the energy splitting for a slab thickness of 20 nm and for four selected gap distances.  For energies $\epsilon_n^\perp$ around the Fermi level, the splitting as a function of energy shows an exponential behavior characteristic for tunnel coupling.  As the slab wavefunction scales with the slab thickness $L$ approximately as $\varphi_n^\perp\sim 1/\sqrt L$, the splitting is proportional to $1/L$.  The open circles in Fig.~2(b) show simulation results for a slab thickness of 10 nm where the energy splitting is reduced by a factor of two, which are in good agreement with the results for the thicker slab.  This demonstrates that the thickness of the simulated slabs is sufficient and quantum confinement effects do not play an important role.  We emphasize that each Kohn-Sham energy $\epsilon_n^\perp$ comes along with a subband of $\bm k_\|$ states, associated with the electron motion parallel to the jellium surface, which is filled up to the Fermi energy.  For this reason, the different Kohn-Sham states contribute differently to the total density $n(x)=\sum_{n\in{\rm occ}}\frac 1\pi(E_F-\epsilon_n^\perp)\left|\varphi_n^\perp(x)\right|^2$ as well as to other quantities such as the conductivity of Eq.~\eqref{eq:sigt}.  

\begin{figure}
\centerline{\includegraphics[width=0.9\columnwidth]{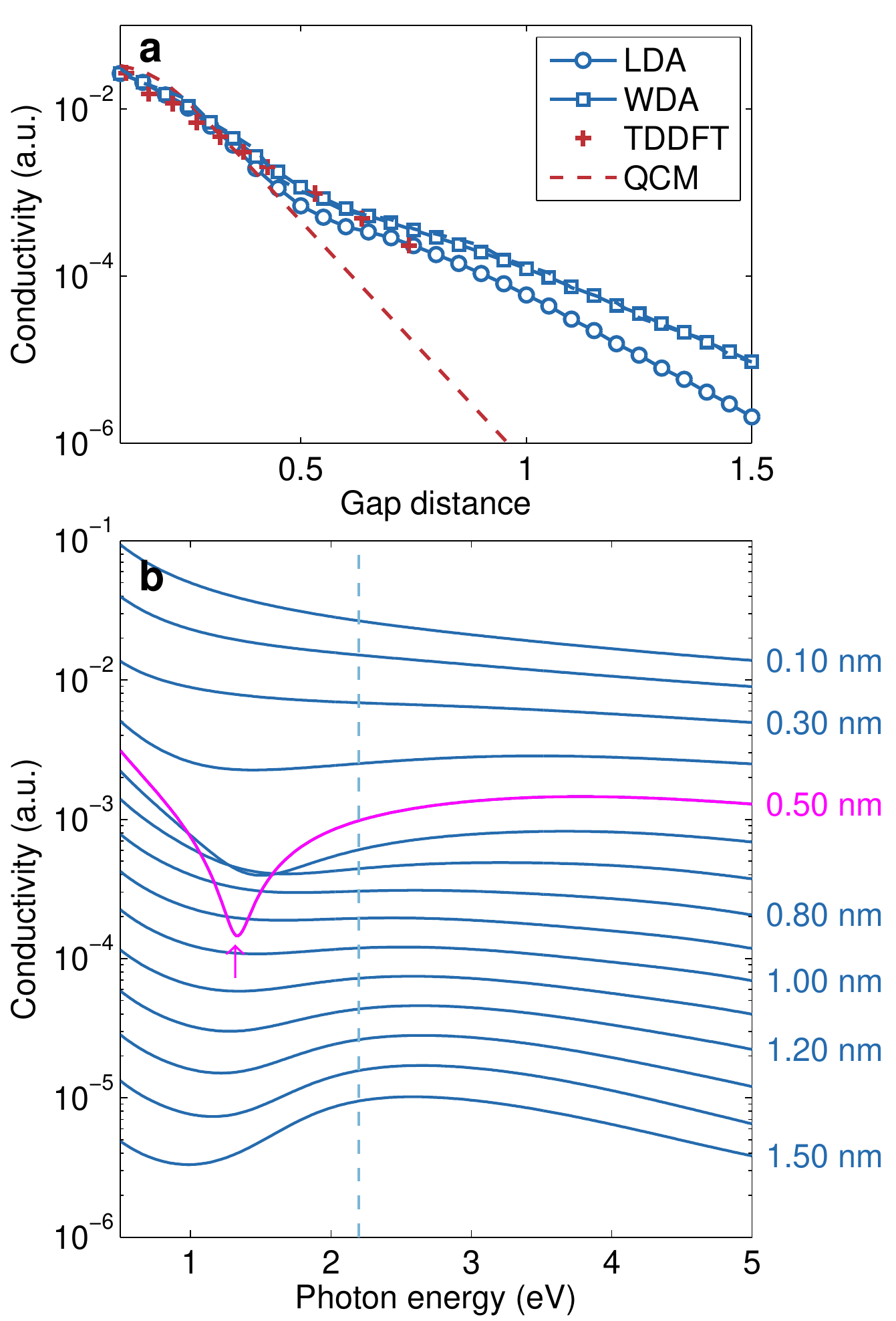}}
\caption{(a) Tunnel conductivity $\sigma_t$ as computed from Eq.~\eqref{eq:sigt} within the LDA (circles) and WDA (square) approximations, and TDDFT and QCM results of Esteban et al.~\cite{esteban:12,esteban:15} (to facilitate the comparison we show $\sigma_t$ in atomic units).  For the QCM expression we use $\omega_p=5.16$ eV, $\gamma_p=0.218$ eV, and $\ell_c=0.075$ nm.  (b) $\sigma_t(\omega,\ell)$ as a function of photon energy $\hbar\omega$ and for selected gap separations $\ell$.
}
\end{figure}

\subsection{DFT tunnel conductivity}  

Fig.~3 shows the tunnel conductivity as computed from Eq.~\eqref{eq:sigt} for a damping constant of $\hbar\eta=0.15$ meV (we checked that somewhat larger or smaller values did not affect the results), and a photon energy of $\hbar\omega=2.2$ eV.  The dashed line shows the results as computed from Eq.~\eqref{eq:drudegap} and the symbols the TDDFT simulation results of Esteban et al.~\cite{esteban:12,esteban:15}.  The square symbols in Fig.~3(a) show WDA results obtained with (solid line) and without (dashed line) periodic boundary conditions which are almost indistinguishable.  For small gap distances all approaches give a very similar decay characteristics, whereas for larger distances, say beyond $\ell=0.5$ nm, our DFT simulations show a significantly slower decay than the predictions of Eq.~\eqref{eq:drudegap}, a finding in agreement with the TDDFT results.  Before pondering on the reasons for this bi-exponential decay, in Fig.~3(b) we show the conductivity as a function of photon energy for a variety of gap distances $\ell$.  One observes a strong frequency dependence of $\sigma_t$, in particular around $\ell=0.5$ nm, in contrast to Eq.~\eqref{eq:drudegap} which predicts for larger gap distances a flat and almost frequency-independent functional dependence of $\sigma_t(\omega,\ell)$.  From the comparison of the WDA and LDA results we observe a relatively small image charge effect, a finding in contrast to the conclusions of Refs.~\cite{esteban:12,esteban:15}.

\begin{figure}
\centerline{\includegraphics[width=\columnwidth]{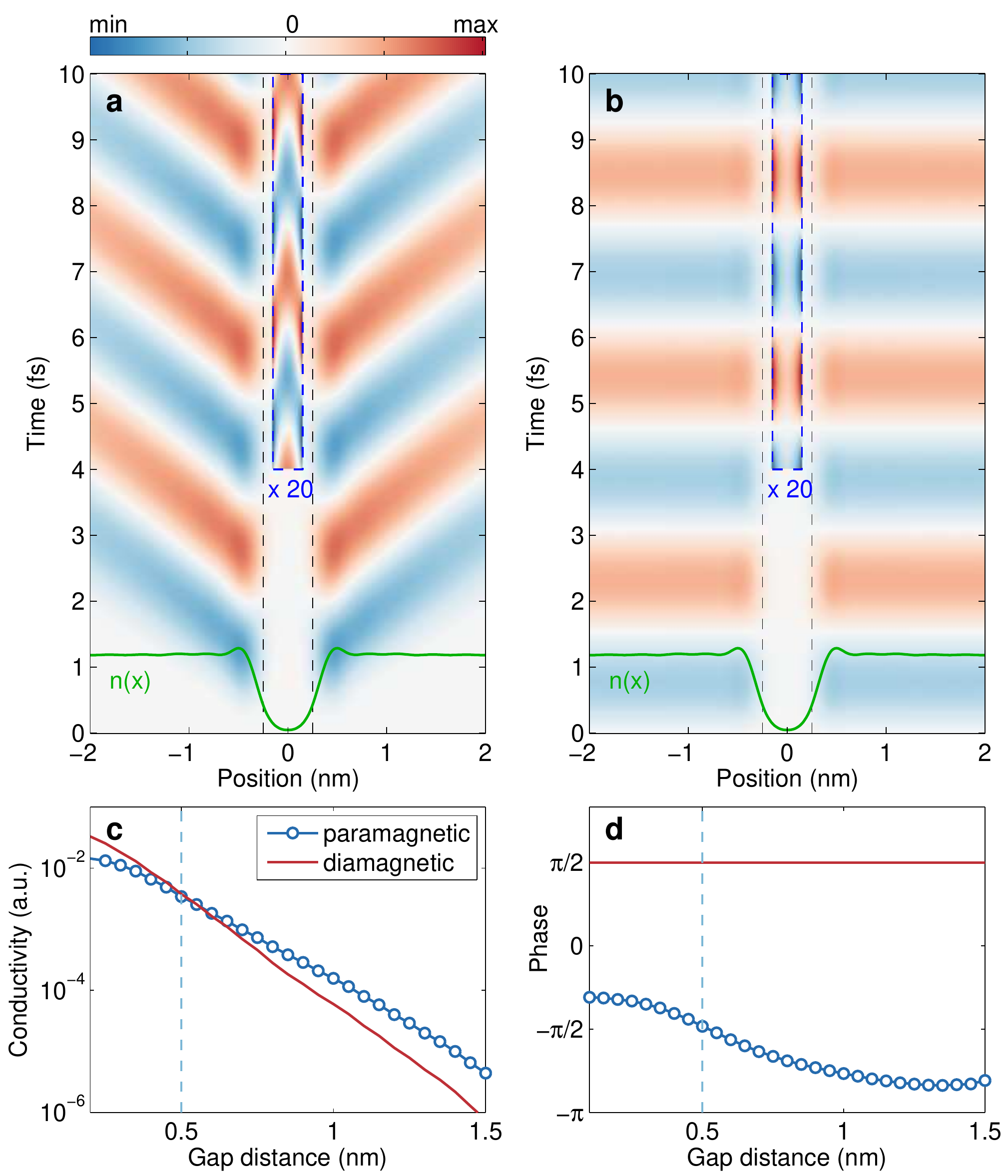}}
\caption{Time dependence of (a) paramagnetic and (b) diamagnetic current for a photon energy of 1.34 eV [see arrow in Fig.~3(b)] and for an external excitation that starts at time zero.  The currents in the gap regions are magnified for clarity by a factor of 20.   (c) Amplitude (square modulus of $\sigma_t$) and (d) phase of paramagnetic (circles) and diamagnetic (solid lines) current as a function of gap distance.}
\end{figure}

To understand the origin of the strong frequency dependence of $\sigma_t(\omega,\ell\approx 0.5\,\mbox{nm})$, in Fig.~4 we show the (a) paramagnetic and (b) diamagnetic current distribution for an optical excitation with $\hbar\omega=1.34$ eV [see arrow in Fig.~3(b)] which is turned on at time zero.  First, the diamagnetic current contribution of panel (b) is proportional to the density $n(x)$ and is phase delayed by 90$^\circ$ with respect to the driving electric field, as can be directly inferred from the second term in Eq.~\eqref{eq:sigt}.  In contrast, the paramagnetic current shown in panel (a) accounts for the creation of current at different positions $x'$ which then flows to position $x$, as described by the first term in Eq.~\eqref{eq:sigt}.  Owing to the $\bm A\cdot\bm p$ light-matter coupling, the highest current contributions originate from regions where the slope of the wavefunction is large, i.e., at the jellium edges, as can be clearly seen in the figure.  In a consecutive step, current flows to different locations where it arrives with some phase delay due to the finite electron velocity.  This is seen most clearly in the gap region, where the current magnitude has been enlarged by a factor of 20 for clarity.

\begin{figure*}
\centerline{\includegraphics[width=1.8\columnwidth]{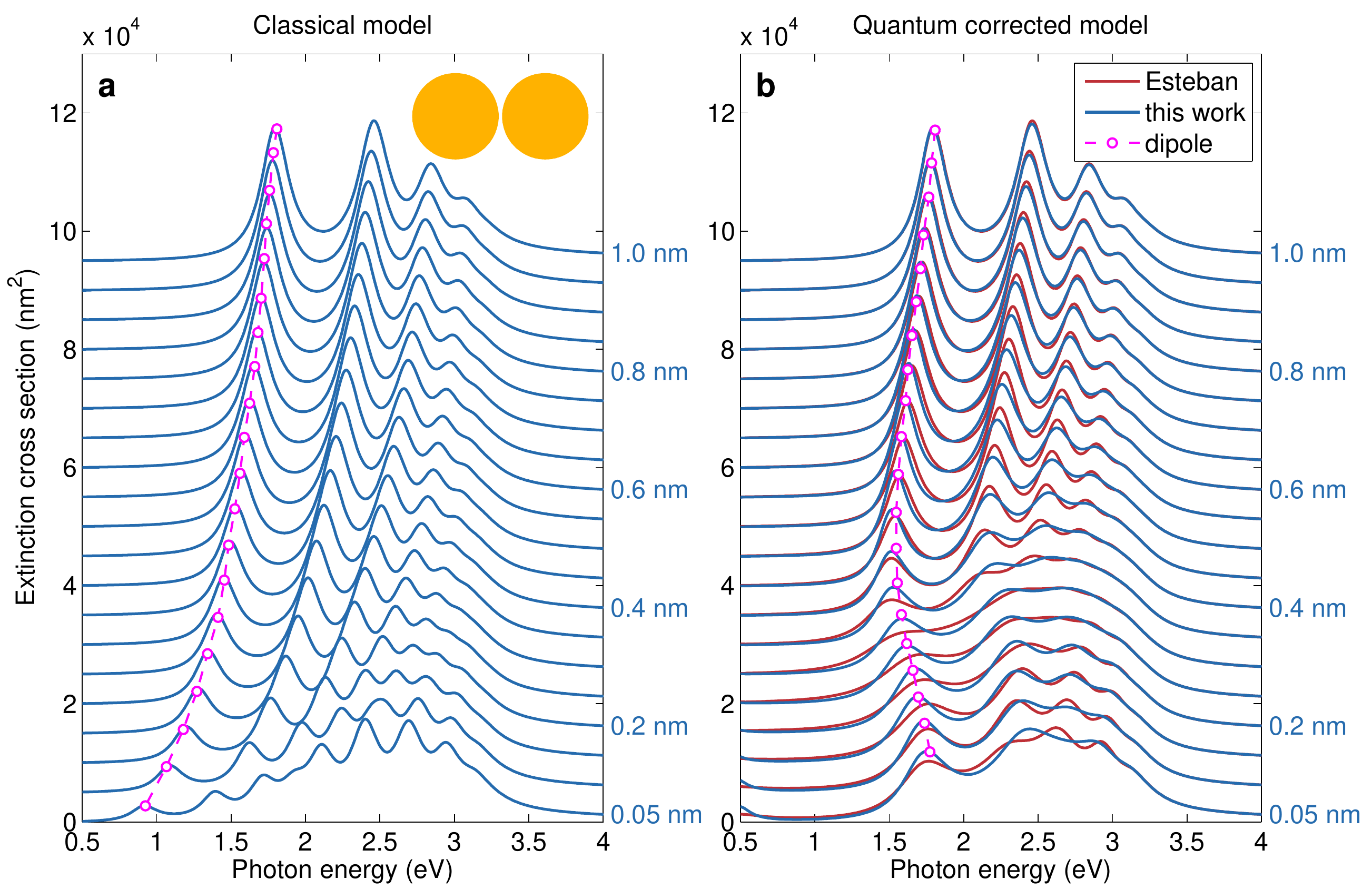}}
\caption{Optical extinction spectra for two coupled nanospheres (50 nm diameter) using the dielectric function for a jellium model representative for sodium, and for electrodynamic simulations (a) without and (b) with consideration of quantum tunneling.  Spectra for different gap separations given on the right-hand side of each panel are offset for better visibility.  The tunnel results are obtained within the boundary QCM approach~\cite{hohenester.prb:15} using the tunnel conductivities of Esteban et al.~\cite{esteban:12,esteban:15} (red lines) and of Eq.~\eqref{eq:sigt}.  Circles indicate the position of the dipole peak, for sodium the charge transfer peak is at photon energies below 0.5 eV.
}
\end{figure*}

We are now in the position to understand the strong frequency and gap distance dependence of $\sigma_t$.  In Fig.~4 we show the (c) amplitude and (d) phase of the paramagnetic and diamagnetic current contributions.  Because we neglect damping effects in the diamagnetic term of Eq.~\eqref{eq:sigt}, an approximation certainly justified for photon energies larger than $\gamma_p$ [see Eq.~\eqref{eq:drudegap} and discussion below], the phase is 90$^\circ$ throughout.  In contrast, due to the charge transport of the paramagnetic term the corresponding current distribution acquires a phase that increases with increasing gap distance (recall that current is predominantly created at the jellium edges).  When the paramagnetic and diamagnetic contributions are out of phase, which happens at a gap distance of approximately 0.5 nm, there is an ideal compensation of these two contributions and $\sigma_t$ exhibits a minimum.  Also the bi-exponential decay shown in Fig.~3(a) can be interpreted as a transition from dominant diamagnetic current at small gap distances to paramagnetic current at large gap distances.

In Fig.~5 we use the tunnel conductivity of Eq.~\eqref{eq:sigt} to compute the optical spectra for two coupled sodium nanospheres, using the quantum corrected model~\cite{esteban:12,esteban:15,hohenester.prb:15}.  Although the $\sigma_t(\omega,\ell)$ values of Eqs.~\eqref{eq:drudegap} and \eqref{eq:sigt} are fairly different, the trends in the computed spectra are similar, showing with decreasing gap distance the emergence of a CTP (gradually appearing at the lowest photon energies and the smallest gap distances) and the blue shift and damping of the bonding plasmonic resonances.  This is finding is in accordance to Refs.~\cite{esteban:15b,knebl:16} where it was shown that $\sigma_t$ mainly triggeres the emergence of the CTP:  below a critical value the CTP peak is absent, above the critical value the peak appears but its spectral position and shape is not substantially influenced by the precise magnitude of $\sigma_t$.

%%%%%%%%%%%%%%%%%%%%
%%%  Discussion  %%%
%%%%%%%%%%%%%%%%%%%%

\section{Discussion}\label{sec:discussion}

The above results finally allow us to critically examine the validity of Eq.~\eqref{eq:drudegap}.  For small gap distances both the QCM of Esteban et al.~\cite{esteban:12,esteban:15} and the DFT results reproduce the bulk conductivity.  We have refrained from introducing a Drude damping for the diamagnetic term in Eq.~\eqref{eq:sigt}, which becomes important in particular for small frequencies, as we expect that such damping should become modified for larger gap distances where the electron density $n(x)$ decreases.  For this reason, all our results are only shown for photon energies above 0.5 eV where our damping neglect is certainly justified.  With increasing gap distance, the QCM and DFT conductivities differ in various ways.  First, the interplay of paramagnetic and diamagnetic currents can lead to interference effects which are absent in the QCM.  In addition, for larger photon energies electron tunneling can occur through excited states [see paramagnetic term in Eq.~\eqref{eq:sigt}] which are more strongly tunnel coupled [see Fig.~2(b)], a feature not included in the QCM expression of  Eq.~\eqref{eq:drudegap}.  

However, even without referring to DFT simulations there are features in the interpolating expression of Eq.~\eqref{eq:drudegap} that call for an improvement.  For large gap distances the conductivity decays as $\sigma_t\sim i(n_0/\gamma_p)e^{-\ell/\ell_c}$, with $n_0$ being the jellium density:  While the exponential dependence of a tunnel process is properly reproduced, the amplitude depends on the damping rate $\gamma_p$ of the bulk material, a finding in marked contrast to STM theory which predicts a conductivity dominated by the tunnel coupling strength.  Additionally, for finite frequencies and small gap distances where the inequality $\omega\gg\gamma_g(\ell)$ holds, Eq.~\eqref{eq:drudegap} predicts an almost gap-distance independent conductivity, contrary to Eq.~\eqref{eq:sigt} where the diamagnetic term (which governs $\sigma_t$ for small distances) decays exponentially due to the vanishing density $n(x)$ in the gap region.  All these findings indicate that the expression of Eq.~\eqref{eq:drudegap} can at best serve as an approximate interpolation function.

In the present work we have investigated sodium which can be modeled reasonably well with the jellium model~\cite{esteban:12,esteban:15}.  For transition metals, such as gold or silver, which are of more importance to the field of plasmonics, one must additionally consider $d$-band contributions, as discussed for instance in Ref.~\cite{esteban:12,esteban:15}.  Because of the larger work functions of gold and silver the gap distances where tunneling plays a role are significantly smaller than for sodium~\cite{zhu:16}.  Whether the interplay of dia and paramagnetic currents, as discussed in this paper, is then still of importance remains to be investigated.

Another open issue that should be addressed in the future is whether there exist quantities besides the optical cross sections that are more strongly influenced by $\sigma_t$.  As mentioned before, the gap conductivity mainly acts as a trigger for the appearence of the CTP peak~\cite{esteban:15b,knebl:16}, however, the precise $\sigma_t$ value is of only minor importance.  Other quantities, such as the amount of charge transferred across the gap, could be influenced more strongly by $\sigma_t$.  Beyond the linear response regime discussed here, nonlinear optical processes, such as higher-harmonic generation, are known to depend sensitively on the details of tunneling~\cite{stolz:14,scalora:14,hajisalem:14} and it might be interesting to extend our approach to this non-linear regime.

\section*{Acknowledgments}

This work has been supported in part by the Austrian science fund FWF under the SFB F49 NextLite and NAWI Graz.  U.H. gratefully acknowledges financial support from the German Science Foundation (DFG), Collaborative Research Project HIOS, SFB 951, for a sabbatical stay at the Humboldt-Universit\"at zu Berlin where part of this work has been performed.  We thank Jorge Sofo and Santiago Rigamonti for most helpful discussions.

%\bibliography{../longbib}

\end{document}